**Experimental Realization of a Passive GHz Frequency-Division Demultiplexer for Magnonic Logic Networks**

*Frank Heussner*, Giacomo Talmelli, Moritz Geilen, Björn Heinz, Thomas Brächer, Thomas Meyer, Florin Ciubotaru, Christoph Adelmann, Kei Yamamoto, Alexander A. Serga, Burkard Hillebrands and Philipp Pirro*

Dr. F. Heussner, M. Geilen, B. Heinz, Dr. T. Brächer, Dr. T. Meyer, Dr. A. A. Serga, Prof. B. Hillebrands, Dr. P. Pirro
Fachbereich Physik and Landesforschungszentrum OPTIMAS
Technische Universität Kaiserslautern
D-67663 Kaiserslautern, Germany
E-mail: heussner@rhrk.uni-kl.de

G. Talmelli, Dr. F. Ciubotaru, Dr. C. Adelmann
Imec
3001 Leuven, Belgium

G. Talmelli
Departement Materiaalkunde, SIEM
KU Leuven
3001 Leuven, Belgium

B. Heinz
Graduate School Materials Science in Mainz
55128 Mainz, Germany

Dr. K. Yamamoto
Advanced Science Research Center
Japan Atomic Energy Agency
Tokai 319-1195, Japan

Dr. T. Meyer
Current affiliation: THATec Innovation GmbH
68165 Mannheim, Germany

Abstract text:

The emerging field of magnonics employs spin waves and their quanta, magnons, to implement wave-based computing on the micro- and nanoscale. Multi-frequency magnon networks would allow for parallel data processing within single logic elements whereas this is not the case with conventional transistor-based electronic logic. However, a lack of experimentally proven solutions to efficiently combine and separate magnons of different frequencies has impeded the intensive use of this concept. In this Letter, the experimental realization of a spin-wave demultiplexer enabling frequency-dependent separation of magnonic signals in the GHz range is demonstrated. The device is based on two-dimensional magnon transport in the form of spin-wave beams in unpatterned magnetic films. The intrinsic frequency-dependence of the beam direction is exploited to realize a passive functioning obviating an external control and additional power consumption. This approach paves the way to magnonic multiplexing circuits enabling simultaneous information transport and processing.

Main text:

The concept of parallel data processing in single elements is very promising since it can multiply the throughput of future logic networks significantly. [1] One important requirement, the wave-based processing of data, has already been demonstrated by utilizing spin waves [2] in micro- and nanostructures. [3, 4] Furthermore, the likewise necessary technique of simultaneous information transport in separated frequency-channels, called frequency-division multiplexing, [5] has been used in various other wave-based applications such as in fiber-optic networks. However, its implementation into magnonic networks has only been studied theoretically [1, 6] or discussed in preparatory works [7, 8, 9, 10, 11] since a device for the efficient combination and separation of spin-wave signals of different frequencies is missing until now. In this work, we present the experimental realization of this required device, which employs caustic-like spin-wave beams to create a so-called magnonic frequency-division demultiplexer. Hence, our results enable the realization of magnonic multi-frequency circuits [6] which allow for parallel data processing in single devices. [1]

Spin waves offer many advantages for wave-based computing [12, 13, 14, 15, 16, 17] due to their micro- to nanometer small wavelengths at GHz-frequencies, [18, 19, 20] their enormous tunability by various parameters [21] and the possibility of charge-less information transport. One exceptional property utilized in the following work is the anisotropic propagation of spin waves in suitable magnetic media, which can lead to the creation of narrow spin-wave beams and caustics [6, 22, 23, 24, 25, 26, 27, 28, 29] (see Experimental Section). Their sub-wavelength transverse aperture [26] is an outstanding reason for their potential use in nanostructured networks, especially when considering the beam formation of high-wavevector magnons. [23] The caustic-like spin-wave beams can be radiated from point-like sources into unpatterned magnetic films [24, 25, 26, 27] and they are formed due to the non-collinearity of the wavevector and the group velocity vector. Furthermore, their propagation direction can be versatilely controlled by, e.g., frequency [6, 22, 24, 27, 28, 29], magnetic field strength [22] and direction of the local magnetization. [25, 26]

In this work, we experimentally demonstrate how these caustic-like spin-wave beams can be used to realize controllable information transport in two-dimensional magnetic films with thicknesses of a few tens of nanometers. The guidance of the signal is not achieved by geometrical structuring [3, 2] or magnetically induced channels [30] but by the inherent anisotropy of the magnonic system. Utilizing this anisotropic signal propagation allows for the realization of passive elements without any additional energy consumption since the magnonic system inherently collimates and steers the energy transport. As just one of the many possible applications of this concept, we present the realization of a passive demultiplexer exploiting the frequency-dependence of the beam direction. By inverting the geometry, a multiplexer can be realized equally well. [31]

The experimental realization of the frequency-division demultiplexer is based on a 30-nm-thin film of the magnetic alloy CoFeB. In contrast to magnetic insulators like yttrium iron garnet (YIG), [28] the metallic ferromagnet CoFeB is much more compatible with state-of-the-art fabrication techniques of silicon-based microchips. Furthermore, CoFeB exhibits a much higher saturation magnetization (see Experimental Section) which enables the creation of caustic-like spin-wave beams in a broad frequency range. Being based on our previous studies, [6] we have designed a demultiplexer prototype by micromagnetic modelling. Subsequently, the sample has been fabricated and studied by detecting the spin-wave intensity using micro-focused Brillouin light scattering spectroscopy (µBLS). [32] After investigating the frequency-dependence of the spin-wave beam directions and comparing it to theoretical calculations, we verify the full functionality of the demultiplexer by two-dimensional measurements of the spin-wave intensity.

**Figure** 1a,b presents the simulated intensity distribution inside of the designed structure for spin waves of frequencies $f_1 = 11.2$ GHz and $f_2 = 13.8$ GHz. The dashed arrows mark the beam directions that are predicted by the developed theoretical model (see Figure 1c and Experimental Section). The functional part of the device comprises the unpatterned area in the center of the structure in which the two-dimensional energy transport takes place. Input and output waveguides, which can serve as interfaces to adjacent magnonic building blocks in larger magnonic networks, are connected to this area. To provide an input signal, magnon excitation is simulated by applying localized alternating magnetic fields $h_{rf}$ inside the input waveguide. The field distribution is chosen in accordance with the experiment in which these fields are created by microwave currents flowing through a microstrip antenna placed across the waveguide. The spin waves propagate towards the unpatterned area and, in agreement with previous observations, [6, 24, 25, 26] two symmetric spin-wave beams are emitted from the waveguide opening, which acts as a point-like source exhibiting a broad wavevector spectrum. The anisotropy of the spin-wave dispersion leads to a strong focusing of the energy. Furthermore, the initial width of the spin-wave beams is determined by the width of the waveguide opening, for which reason an abrupt transition into the unstructured film area without widening was chosen. The observed frequency dependence of the beam directions is utilized to realize the demultiplexing functionality by properly adding two output waveguides after a certain propagation distance. Their positions are asymmetric with respect to the input, leading to the following consequence: at frequency $f_1 = 11.2$ GHz (Figure 1a), the upper spin-wave beam, which propagates at an angle of $\theta_{\mathrm{B},1}^{\mathrm{theo}} = 77.6^{\circ}$ as predicted by the theoretical model, is channelled into output 1 and transmits the information through the device. The lower one is blocked by the edge of the magnetic structure so that the signal cannot reach output 2. In contrast, the beam angles of spin waves of frequency $f_2 = 13.8$ GHz are changed to $\theta_{\mathrm{B},2}^{\mathrm{theo}} = 68.9^{\circ}$ ($\theta_{\mathrm{B},2}^{'} = 180^{\circ} - \theta_{\mathrm{B},2}$, resp.) resulting in a sole signal transmission into output 2 (Figure 1b). Hence, the device separates spin-wave signals of two frequencies into two spatially detached output waveguides, whose transition zones are especially tailored to enable an efficient channelling of the spin waves and to avoid reflections. The exploited frequency dependence of the beam directions results from the modification of the spin-wave isofrequency curve with frequency (see Figure 1c) and the according variation of the group-velocity vectors, which are always perpendicular to this curve (see Experimental Section). It should be mentioned that the exact operating frequencies of the device can be controlled by changing, e.g., the external field strength since this shifts the magnon dispersion. Here, the external magnetic field is set to $\mu_0 H_{\mathrm{ext}} = 75$ mT.

Based on the design developed and optimized by micromagnetic modelling, the prototype of the passive demultiplexer has been fabricated based on a 30-nm-thin film of CoFeB as shown in Figure 1d. To prove the functionality of the frequency-division demultiplexer, μBLS measurements of the spin-wave intensity have been performed at two positions in the center of the output waveguides as marked in Figure 1d. Spin waves are excited in the input in a frequency range from 10.5 GHz to 15.5 GHz in steps of 0.1 GHz by applying an according microwave signal to the microstrip antenna. **Figure** 2a shows the detected frequency dependence of the spin-wave intensity in the two spatially detached output waveguides. A clear separation of spin-wave signals depending on their frequency is visible. Signals in the narrow range from 10.8 GHz to 11.8 GHz are channeled into output 1. In contrast, output 2 accepts only spin waves in the broader frequency band from 12.3 GHz to 15.3 GHz, hence, without any overlap. This measurement clearly demonstrates that spin-wave signals with appropriately chosen frequencies, e.g., the frequencies of maximum intensity in the different outputs $f_1 = 11.2$ GHz and $f_2 = 13.8$ GHz, are spatially separated by the prototype device as predicted by the micromagnetic modelling.

The beam angle $\theta_{\mathrm{B}}$ shown in Figure 2b is experimentally determined by measuring the spin-wave intensity along a line in the input transition zone (left green dotted line, Figure 1d) and along a line in the second third of the unpatterned area (right green dotted line). From the observed shift of the intensity maxima, the angle $\theta_{\mathrm{B}}$ is calculated. No beams are detected below 10.8 GHz since the dispersion relation is very flat until the frequency of ferromagnetic resonance $f_{\mathrm{FMR}} = 10.94$ GHz is reached. The measured data agrees well with the theoretically predicted curve (solid line in Figure 2b). This is due to the fact that, in addition to the peculiarities of anisotropic magnon propagation, [22] our model considers such important parameters like the excited wavevector spectrum and the distance between the observation point and the source. The optimum angles $\theta_{\mathrm{B},1}^{\mathrm{exp}} = 77.9^{\circ}$ and $\theta_{\mathrm{B},2}^{\mathrm{exp}} = 69.6^{\circ}$, leading to maximum intensity in the outputs, are indicated in Figure 2b together with the shaded acceptance intervals, which result from the assumption that beams are channeled into the outputs if their direction deviates less than 2.5° from the optimum angles. These acceptance intervals in combination with the curve $\theta_{\mathrm{B}}(f)$ can explain the differing widths of the accepted frequency bands of the two outputs: The angle variation is quite large in the region of $f_1 = 11.2$ GHz. Hence, only frequencies within a small range around $f_1$ create beams whose directions lie inside the angular acceptance interval of output 1. In contrast, the slope of the mentioned curve is significantly lower at $f_2 = 13.8$ GHz. Thus, a larger variation of the frequency is permitted around $f_2$ until the occurring beam angles are out of the acceptance interval of output 2.

For a detailed visualization, the spin-wave intensity was measured in the entire demultiplexer area at the frequencies $f_1$ and $f_2$ by performing two-dimensional μBLS scans. (The experimental proof of the simultaneous transmission of spin waves of these two frequencies through a shared waveguide can be found in the Supporting Information.) A comparison of the two-dimensional measurements (**Figure 3**) with the results of the micromagnetic modelling (Figure 1a,b) reveals a remarkably good agreement. This confirms that the whole demultiplexing mechanism works exactly as predicted and is based on the creation of spin-wave beams, their frequency-dependent propagation though the unstructured area, and the channeling of the signals into different outputs.

Furthermore, these measurements can be used to determine the input-output ratio $\kappa$ of the device to reveal the influence of the beam formation on signal losses. Considering the transmission from the input waveguide into output 1 at frequency $f_1$ yields a value of $\kappa_1^{\mathrm{exp}} = (-18.1 \pm 2.1)$ dB whereas a ratio of $\kappa_2^{\mathrm{exp}} = (-18.2 \pm 1.9)$ dB is obtained for a signal at frequency $f_2$ channeled into output 2. For comparison, the ratio $\kappa^{\mathrm{theo}}$ can be calculated analytically considering the energy splitting and the propagation losses only (see Experimental Section). The resulting values $\kappa_1^{\mathrm{theo}} = (-18.95 \pm 1.84)$ dB and $\kappa_2^{\mathrm{theo}} = (-17.64 \pm 1.69)$ dB show that the major fraction of the losses arise from the intrinsic spin-wave damping during their propagation and not due the demultiplexing mechanism based on beam formation and channeling (the splitting is suppressible [6]).

Finally, the scalability and advantages of the presented concept should be discussed. The employed spin-wave beams are created due to the anisotropy of the magnonic system and, for limited length scales, their width is primary determined by the size of the source. [26] In the presented case, the anisotropy is induced by the dipole-dipole-interaction, which is dominant in the low-wavevector range of the magnon spectrum. However, a beam formation is also possible for high-wavevector spin waves whose dispersion characteristics are mainly dominated by the exchange interaction. [23] Hence, the presented concept to guide two-dimensional energy transport by spin-wave beams, formed due to the intrinsic anisotropy of the system, can be scaled down to significantly smaller length scales. Furthermore, the passive character of the signal steering is an important advantage over other concepts requiring energy-consuming external control. This is the case, e.g., with the previously demonstrated spin-wave time-division multiplexer, which relies on external charge currents. [33]

In conclusion, we have developed and experimentally realized a passive frequency-division demultiplexer, which is based on intrinsically steered two-dimensional signal transport in unpatterned magnetic films. The prototype device enables the spatial separation of spin-wave signals of different frequencies. It exploits the frequency-dependence of the direction of narrow spin-wave beams, which are formed in in-plane magnetized films due to the anisotropic spin-wave dispersion. A theoretical approach to predict the beam directions has been developed and verified. The utilized beam steering requires no external control or additional power consumption and it can be used to realize a whole multi-frequency circuit which enables simultaneous data transport through single transmission lines by employing the technique of frequency-division multiplexing. This is the basis for exploiting one of the main advantages of wave-based computing, namely the technique of parallel data processing in single devices, which can now be implemented in an energy-efficient way building on the presented prototype.

**Experimental Section**

*Micro-focused Brillouin light scattering (μBLS) measurements:* The spin-wave intensity is experimentally measured by utilizing μBLS [32] with a spatial resolution of around 250 nm. In all cases, the shown spin-wave intensity of frequency $f$ is the detected time-averaged μBLS intensity integrated over the frequency interval [$f$ - 0.25 GHz, $f$ + 0.25 GHz]. The error of the measured beam angles $\theta_B$ shown in Figure 2b results from the spatial resolution of the setup and is calculated to 3.5°. For reasons of clarity, error bars are omitted in the figure. The spin-wave intensity shown in Figure 3 is normalized along the $y$-direction to the interval [0,1] to compensate for the background noise of the setup and for the spin-wave damping so that a clear comparison of the two output signals is easily possible. Due to this normalization, the visibility of spin-wave beams, which are reflected from the edge of the structure and which are much weaker than the incident beams, is reduced.

*Micromagnetic modelling:* The micromagnetic modelling is carried out by using the GPU-accelerated simulation program MuMax3. [34] The simulated structure of the size 31.5 µm x 15 µm x 30 nm is discretized into cells of approximately 15 nm x 15 nm x 30 nm leading to a consideration of in-plane wavevectors of at least up to 100 rad µm$^{-1}$. Furthermore, the experimentally determined material parameters of CoFeB ($M_S$ = 1558 kA m$^{-1}$, $A_{ex}$ = 17.6 pJ m$^{-1}$, $\alpha$ = 0.0043) are used. On the outer vertical edges of the waveguides (shape of the structure as shown in Figure 1), the damping is incrementally increased (25 steps over a distance of 0.5 µm) to a value of $\alpha$ = 0.5 to suppress reflections of the spin-wave energy. Furthermore, periodic boundary conditions are used along the horizontal edges of the unstructured film area. Spin-wave excitation and propagation are simulated for 7.5 ns by applying an alternating magnetic field inside the input waveguide. The used field distribution is calculated according to Biot-Savart's law for a 0.5 µm wide and 100 nm thick microstrip antenna (as used in the experiment) through which a AC current flows (with 5 mA amplitude). The resulting raw data of the modelling is the time-dependent vector of the magnetization inside every cell, which has been saved every 25 ps. A temporal Fourier-transformation of the data from 2.5 ns to 7.5 ns and a squaring of the resulting amplitude has been performed to calculate the frequency-dependent, time-averaged spin-wave intensity. This intensity is integrated over the frequency interval [$f$ - 0.25 GHz, $f$ + 0.25 GHz], so that a direct comparison with the experimental results obtained by μBLS measurements is possible. Furthermore, the simulated spin-wave intensity distribution shown in Figure 1 is normalized along the $y$-direction to its sum to compensate for the spin-wave damping and to allow for a clear comparison of the two output signals.

*Sample fabrication:* The basis of the sample fabrication [35] is a 30-nm-thin film of the magnetic alloy CoFeB, which is deposited as part of the layer stack Ta(3 nm)/CoFeB(30 nm)/Ta(3 nm) onto a Si/$SiO_2$(300 nm) wafer. First, the material parameters of this unpatterned sample have been measured using two methods. The saturation magnetization $M_S$ = 1558 kA $m^{-1}$ and the Gilbert damping constant $\alpha$ = 0.0043 have been determined by measurements of the ferromagnetic resonance (FMR). Furthermore, the exchange constant $A_{ex}$ = 17.6 pJ $m^{-1}$ results from μBLS measurements of the thermal magnon spectrum and a comparison to analytical calculations according to the theory. [36] Afterwards, the magnetic layer is structured by e-beam lithography of a hydrogen silsesquioxane (HSQ) hard mask and subsequent ion beam etching, leaving around 70 nm of the resist on top of the structure. To reduce the height differences, a planarization step using Spin-on Carbon (SoC) follows with a subsequent deposition of a 30-nm-thin layer of SiN (150°C, Chemical Vapor Deposition). Finally, the microstrip antenna is fabricated by creating a mask using polymethylmethacrylat (PMMA) resist and e-beam lithography followed by sputtering deposition of Ti (10 nm)/Au (100 nm) and the removal of the unwanted parts by a lift-off process in acetone. The resulting sample is shown in Figure 1d. The magnetic structure consists of an unpatterned film area and three 1.5 μm wide waveguides. The positions of the output waveguides along the $y$-direction are asymmetric with respect to the input waveguide which enables the demultiplexing functionality of the device as demonstrated in the main text. The light shading on the edge of the transition zones is due to some residual material, which, however, doesn't influence the functionality of the device. The 0.5 μm wide and 100 nm thick microstrip antenna, which is placed on top of the input waveguide, is connected to a microwave setup consisting of a microwave generator and an amplifier which provide the high-frequency current with a nominal output power of $P = +7$ dBm. The final power reaching the sample is very likely significantly reduced. The rectangular structures around the actual device are used to stabilize the positioning of the μBLS microscope. It should be mentioned that the dimensions of the presented prototype have been chosen to allow for an optical investigation using μBLS. Further miniaturization is possible as discussed in the main text.

*Calculation of the input-output ratio κ:* To calculate the experimental values $\kappa^{\mathrm{exp}}$ of the input-output ratio, the spin-wave intensity at the position 0.5 μm in front of the input transition zone and the intensity 5.1 μm behind the output transition zone is integrated over the waveguide width and both results are divided by each other. The mentioned error results from the deviations around the mean value in case the surrounding measuring points are considered. For both frequencies, the detected intensity inside the output, which is not hit by the spin-wave beam, is equal or only marginally above the noise-level of the μBLS setup. This highlights the pronounced demultiplexing functionality.

The theoretical reference values $\kappa^{\mathrm{theo}}$ are calculated from the distances $L$ which the spin waves have to travel from the position in the input to the position in the output, where the intensity has been measured to determine the experimental value. The calculations are carried out according to the formula

$$\kappa^{\mathrm{theo}} = 10 \log_{10}[1/2 \exp(-L_{\mathrm{f}}/(0.5\, v_{\mathrm{f}}\, \tau_{\mathrm{f}}) - L_{\mathrm{w}}/(0.5\, v_{\mathrm{w}}\, \tau_{\mathrm{w}}))],$$

in which the occurring energy splitting into two spin-wave beams is considered by the factor 1/2. It should be mentioned that the splitting can be suppressed by changing the design of the demultiplexer as shown in ref. [6]. Furthermore, the factor 0.5 has to be introduced since the spin-wave intensity is measured instead of the amplitude.

The group velocity $v$ of the spin waves differs inside the waveguides ($v_{\mathrm{w}}$) and in the unpatterned film area ($v_{\mathrm{f}}$) and can be calculated according to ref. [36, 37]. To determine $v_{\mathrm{f}}$, the spin-wave wavevector relating to the occurring beam angle is used for the calculations. Furthermore, $L_{\mathrm{w}}$ is the spin-wave propagation distance in the waveguides whereas $L_{\mathrm{f}}$ is the corresponding distance in the film area. Finally, the spin-wave lifetime $\tau$ results from the damping parameter $\alpha$ and is approximated by the value $\tau(k = 0)$ at ferromagnetic resonance. [37] For all calculations, the effective values of the magnetic field $\mu_0 H_{\mathrm{eff}}$ (75 mT inside the film, 50 mT inside the waveguide) and the waveguide width $w_{\mathrm{eff}} = 1.3$ µm are used. These effective values result from the demagnetising field at the edge of the waveguides and can be extracted from the micromagnetic modelling or calculated as shown in ref. [37].

The following values are obtained for the different frequencies $f_1 = 11.2$ GHz and $f_2 = 13.8$ GHz: $L_{1,\mathrm{w}} = 5.48$ µm, $L_{1,\mathrm{f}} = 13.05$ µm, $v_{1,\mathrm{w}} = 11.65$ µm ns$^{-1}$, $v_{1,\mathrm{f}} = 7.08$ µm ns$^{-1}$, $\tau_{\mathrm{w}} = 1.284$ ns, $\tau_{\mathrm{f}} = 1.254$ ns, and $L_{2,\mathrm{w}} = 5.51$ µm, $L_{2,\mathrm{f}} = 13.90$ µm, $v_{2,\mathrm{w}} = 9.70$ µm ns$^{-1}$, $v_{2,\mathrm{f}} = 8.93$ µm ns$^{-1}$ resulting in the values for the input-output ratio $\kappa^{\mathrm{theo}}$ as mentioned in the main text. The error of $\kappa^{\mathrm{theo}}$ is calculated by assuming a slightly larger Gilbert damping parameter ($\alpha = 0.0048$), which is very likely to occur due to, e.g., the structuring process of the device.

*Theoretical description of spin-wave beams and caustics:* The basis of the developed theoretical model is the anisotropic isofrequency curve (see Figure 1c), which describes the dependence of the wavevector component $k_y$ on the component $k_x$ at a particular frequency $f$. The dispersion calculations are based on ref. [36] taking into account $k_y \parallel H_{\mathrm{ext}}$, $k_x \perp H_{\mathrm{ext}}$, $\mu_0 H_{\mathrm{ext}} = 75$ mT and the material parameters of CoFeB. The direction of the anisotropic spin-wave propagation can be derived from the curve since the spin-wave group velocity vector is perpendicular to it at every point. For a precise prediction of the resulting spin-wave beam orientation, the excited wavevector spectrum $A_k$ and the distance $d$ between the beam source and the observation point is considered as explained in the following. This is the main difference to a previous theoretical approach [22] and our results reveal, that the observed direction of maximum focusing is not always given by the caustic direction (According to ref. [22], caustics occur at the points of the isofrequency curve where its curvature is zero). Due to this reason, we refer to the observed beams resulting from the focused energy transport as (caustic-like) spin-wave beams instead of caustics.

The first step of the developed approach is to include the wavevector spectrum $A_k$ of the beam source in the calculations. It can be assumed that the spin waves propagate through the input in the form of the first waveguide mode having a sinusoidal shape along the short axis of the waveguide and only one maximum in the centre. If the input waves reach the transition zone at the connection between the waveguide and the unstructured area, the waveguide mode acts as a one-dimensional source (with effective width $w_{\mathrm{eff}}$, explanation above) for secondary spin waves propagating into the unpatterned film area and forming beams due to their anisotropic propagation. The confinement of the mode across the waveguide leads to a broad spectrum $A_k$ of the respective wavevector component $k_y$, which can be calculated to $A_k(k_y) = \cos(k_y w_{\mathrm{eff}}/2)/(\pi^2/w_{\mathrm{eff}}^2 - k_y^2)$ by spatial Fourier-transformation of the mode profile. This spectrum is considered by using the following procedure. First, the angle $\theta_{v\mathrm{g}}$ between the group velocity vector and the external field direction is calculated from the isofrequency curve as a function of the wavevector component $k_y$. Second, the wavevector spectrum $A_k(k_y)$ is projected onto the curve $\theta_{v\mathrm{g}}(k_y)$ by a numerical integration to determine the amount of excited spin waves propagating into the direction $\theta_{v\mathrm{g}}$. The result of this procedure is the spin-wave amplitude $A_{\mathrm{SW}}^{\mathrm{initial}}(\theta_{v\mathrm{g}})$, which reflects the angle distribution of the spin-wave flow under consideration of the initial wavevector spectrum $A_k$. Maxima of $A_{\mathrm{SW}}^{\mathrm{initial}}(\theta_{v\mathrm{g}})$ reveal a wave focusing into the respective directions. Their occurrence can be explained by the fact that the anisotropy of the system leads to isofrequency curves with expanded regions where the group velocity direction $\theta_{vg}$ is (nearly) unchanged for a broad range of wavevectors $k_y$. If the excited spectrum $A_k$ has a significant magnitude in these ranges of $k_y$, an amplitude concentration occurs into the respective directions $\theta_{vg}$ leading to the maxima of $A_{\mathrm{SW}}^{\mathrm{initial}}(\theta_{vg})$. Hence, these maxima indicate the beam creation.

However, a second step is necessary to precisely determine the beam directions. The beam source described above has a certain extent which leads to the fact that the overall spin-wave amplitude at the observation point results from a superposition of spin waves which are generated at different positions within the source. If the observation point is close, the source appears under a large angular range and spin waves from considerably different directions superpose. In this case, the focusing pattern can be significantly influenced. This effect is included in the further calculations by considering both the initial angle distribution of the spin-wave amplitudes $A_{\mathrm{SW}}^{\mathrm{initial}}(\theta_{vg})$ and the sinusoidal mode profile in the transition zone, which represents the source of the beams. Based on these quantities, the final spin-wave amplitude $A_{\mathrm{SW}}^{\mathrm{final}}$ at a given observation point is calculated by integrating the amplitudes of all spin waves which originate at different points of the source and reach the observation point from different directions. Finally, this amplitude is squared to obtain the overall spin-wave intensity $I_{\mathrm{SW}}^{\mathrm{final}}(\theta_{vg}, d)$ occurring at a distance $d$ from the source and in the direction $\theta_{vg}$ in relation to the external magnetic field.

Spin-wave beams occur if the curve $I_{\mathrm{SW}}^{\mathrm{final}}(\theta_{vg}, d)$ exhibits pronounced peaks into certain directions $\theta_{\mathrm{B}}$. The angles $\theta_{\mathrm{B}}$ of the theory curve shown in Figure 2b belong to the maxima of the calculated intensity distribution $I_{\mathrm{SW}}^{\mathrm{final}}(\theta_{vg})$ when assuming a distance of $d = 10$ μm and using the parameters of the investigated system (material parameters of CoFeB, $\mu_0 H_{\mathrm{ext}} = 75$ mT, $w_{\mathrm{eff}} = 1.3$ μm, $f \in [10.8, 15.4]$ GHz).

**Acknowledgements**
Financial support by DFG within project B01 and A03 of the Transregional Collaborative Research Centre (SFB/TRR) 173 "Spin+X" and from the EU Horizon 2020 research and innovation programme within the CHIRON project (contract number 801055) is gratefully acknowledged. The sample fabrication has been supported by imec's Industrial Affiliation Program on Beyond-CMOS Logic. B.He. was supported by a fellowship of the Graduate School Materials Science in Mainz (MAINZ) through DFG funding of the Excellence Initiative (GSC-266).

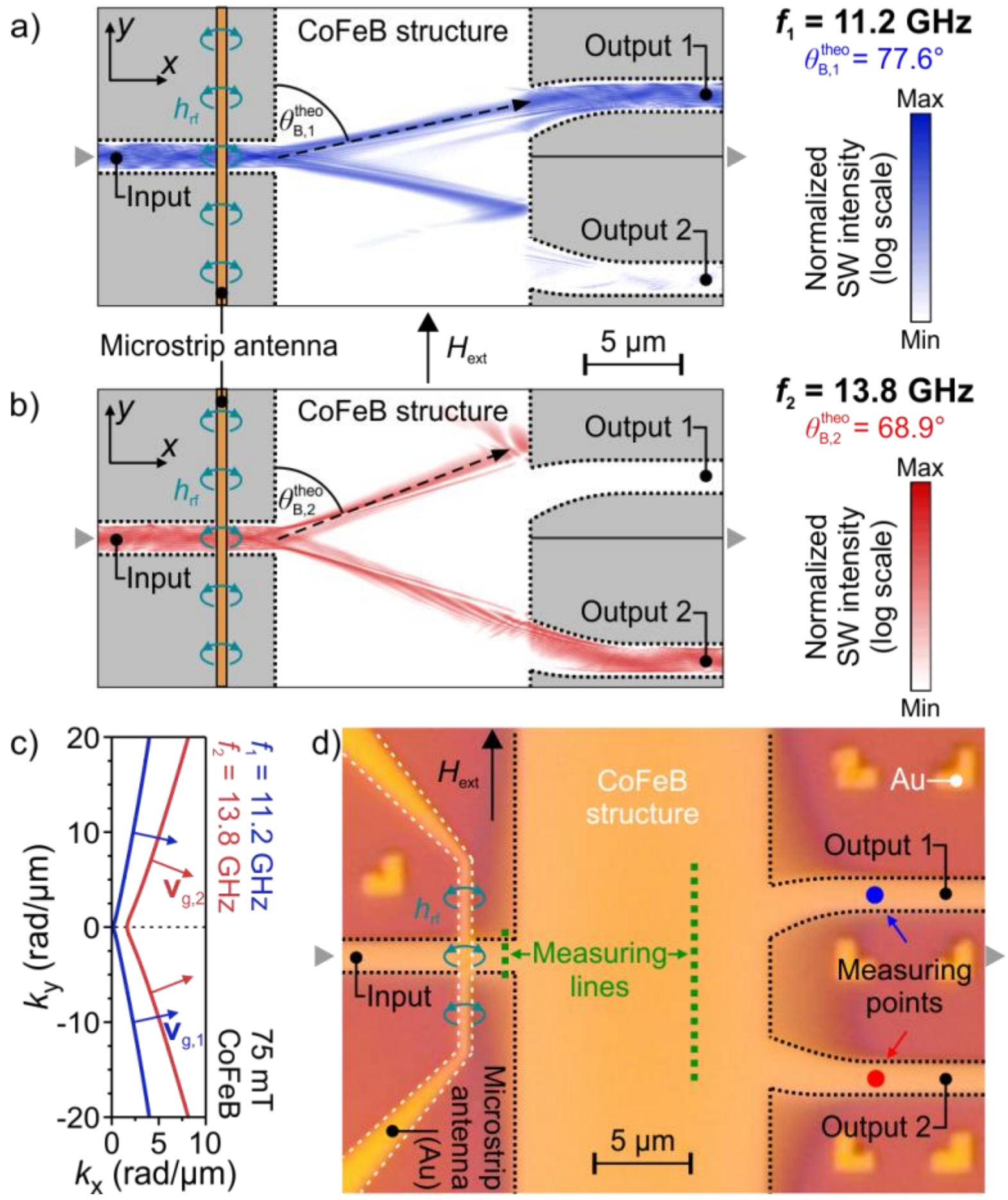

**Figure 1.** Micromagnetic modelling of the frequency-division demultiplexer and microscope image of the fabricated sample. The positions of the output waveguides along the $y$-direction are asymmetric with respect to the input waveguide which enables the demultiplexing functionality of the device. a) Design of the demultiplexer and simulated spin-wave propagation at $f_1 = 11.2$ GHz. The signal is guided from the input waveguide into output 1 by a spin-wave beam. b) Spin-wave distribution at $f_2 = 13.8$ GHz. The signal is channelled into output 2 due to the changed direction of the spin-wave beams in the unstructured film area. c) Isofrequency curves of the studied system. The directions of the group velocity vectors, which are perpendicular to the curve, vary with spin-wave carrier frequency. d) Fabricated sample according to the developed layout of the demultiplexer.

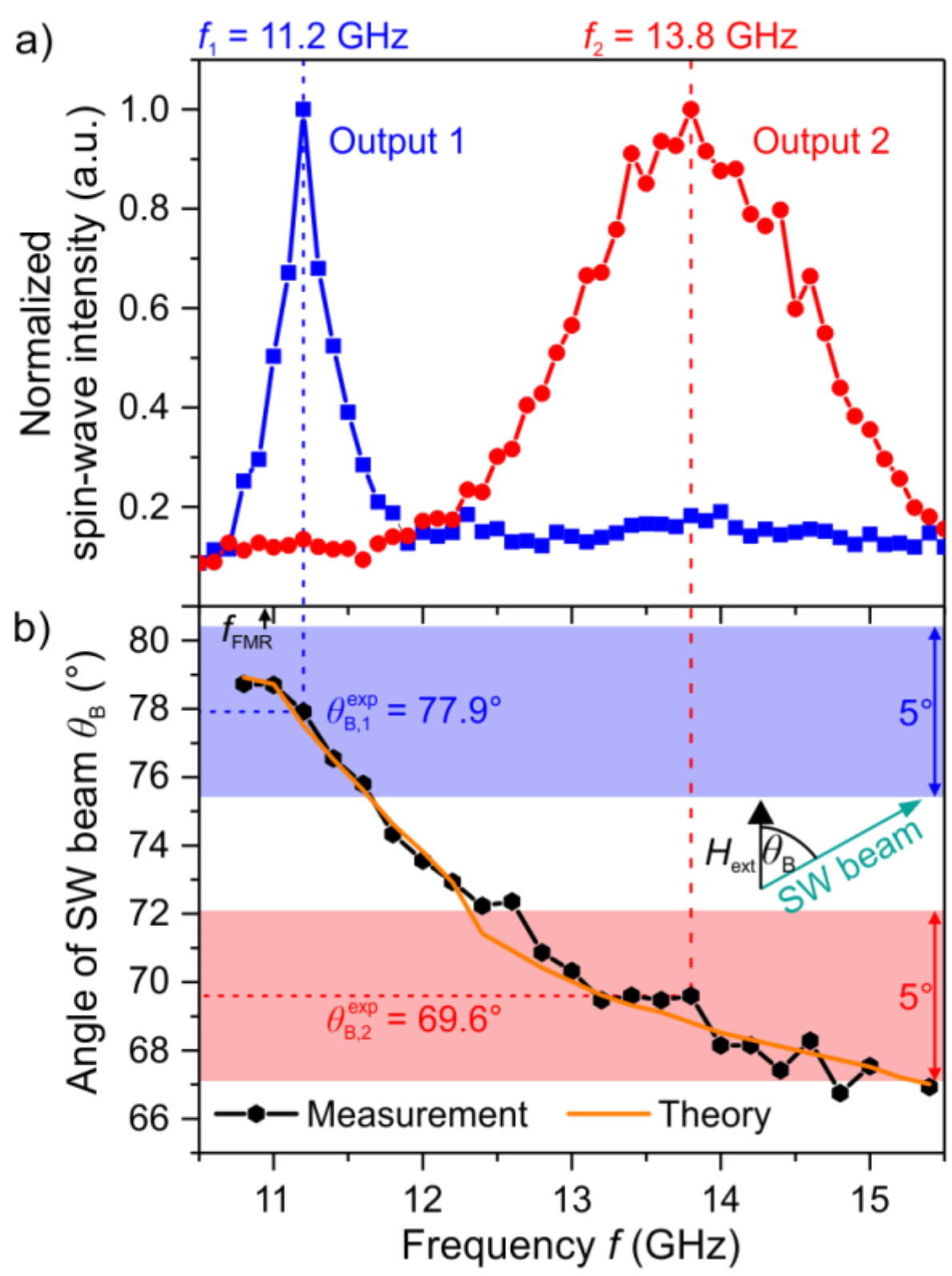

**Figure 2.** Experimental realization of spin-wave demultiplexing using the developed prototype device. a) Frequency-dependence of the spin-wave intensity in the output waveguides measured by µBLS. Spin-wave signals are detected inside the two outputs within non-overlapping frequency ranges exhibiting signal maxima at the frequencies $f_1$ and $f_2$. b) The measured frequency-dependence of the direction of the spin-wave beams is compared with the developed theoretical model. Shaded areas mark the angular intervals of signal acceptance of both outputs. The centres of the intervals are given by the optimal angles $\theta_{B,1}^{exp}$ and $\theta_{B,2}^{exp}$.

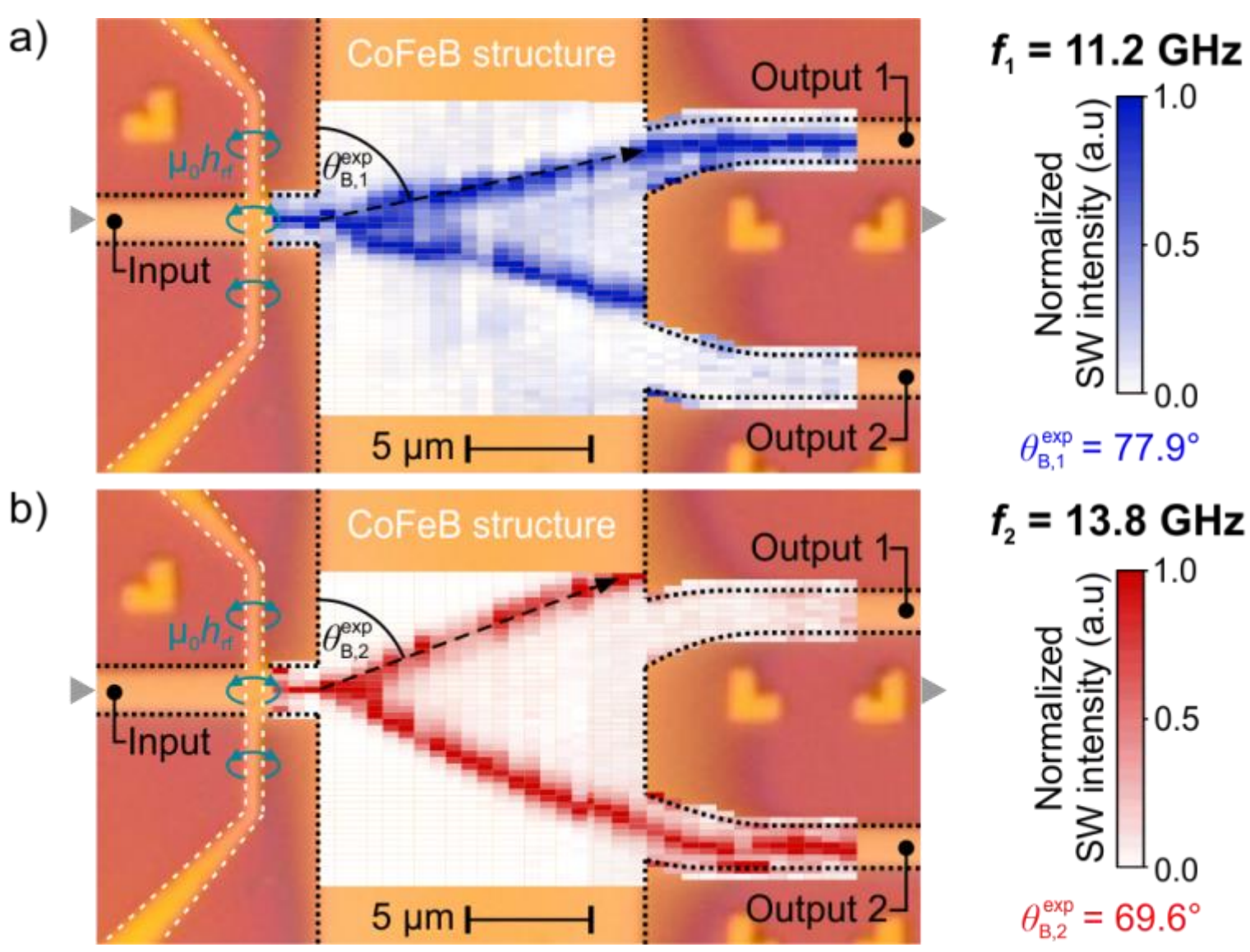

**Figure 3.** Experimental verification of the beam-based demultiplexing mechanism. µBLS measurements of the spin-wave intensity are performed to visualize the two-dimensional spin-wave transport. The separation of spin-wave signals from one shared input waveguide by frequency-dependent channelling into spatially detached output waveguides occurs due to the intrinsically controlled beam formation. a) Spin wave guidance into output 1 at $f_1 = 11.2$ GHz and ,b) into output 2 at $f_2 = 13.8$ GHz.